\documentstyle[12pt]{article}
\begin{document}
\large
\begin{center}
{\bf Some Remarks to the Standard Model of Electroweak
Interactions. Absence of Polarization i.e. Charge Renormalization
and Mass Generations}\\
\vspace{1cm}

Beshtoev Kh. M.\\
\vspace{1cm}
Joint Institute for Nuclear Research, Joliot Curie 6, 141980 Dubna,
Moscow Region, Russia;
Scient. Research Institute of Applied Mathematics and Automation of KBSC RAS,
Shortanova 89a, Nalchik, Kabardino-Balk. Rep., Russia\\

\end{center}

\vspace{0.5cm}
\par
{\bf Abstract}\\

\par
In the article considers some remarks to the standard approach to weak
interactions. Higgs mechanism contains contradictions, therefore cannot
be considered as a realistic mechanism for mass generations.
The couple constant of the weak interactions cannot be changed in
dependence on momenta transfer (it leads to violation of the
law of energy-momenta conversation), i.e. it must be constant. It means
that there are no charge vacuum polarizations or renormalization of the
couple constant of the weak interactions. Then it is clear that the
resonance enhancement of neutrino oscillations in matter cannot
exist either for absence of matter polarizations.
Masses are not generated and connected states cannot
form in these interactions either since $P$-parity is violated.
Then also as in the strong and electromagnetic interactions, in
the weak interactions we can use the perturbative theory but in
this case the propagators must be the propagators of free particles
without renormalization.

\section{Introduction}

At present there are three families of quarks and leptons [1]:
$$
\begin{array}{cc} u& \nu _{e}\\ d& e \end{array} ;\qquad
\begin{array}{cc}c& \nu _{\mu }\\ s& \nu \end{array} ;\qquad
\begin{array}{cc} t& \nu _{\tau}\\b& \tau \end{array} .
\eqno(2.1)
$$
It is supposed that all available experimental data agree well with the
standard model of the electroweak interaction [2] proposed by Glashow,
Weinberg, and Salam for three quark and lepton families.
In the framework of electroweak  interactions, i.e. at $W$
and $Z$ boson exchanges the transitions between different families do not take
place. These transitions are realized beyond the electroweak interactions [1].
\par
In the weak interactions $P$-parity is violated,
in contrast to the electromagnetic and strong interactions.
It is a distinctive feature of the weak
interactions which leads to some consequences. At first we give elements of
the electroweak model and then come to consideration of concrete remarks
concerning these consequences.

\section{Elements of the Standard Model of Electroweak Interactions
and Some Remarks to This Model}

Consider elements of the electroweak interactions model and concrete remarks.\\

\par
2.1. {\bf Elements of the Standard Model of Electroweak Interaction}

 \par
 The Lagrangian of the theory contains left and right doublets of leptons and
 quarks,
\par
$$
\Psi _{lL} = \left(\begin{array}{c} \nu _{l}\\ l\end{array}\right)_L ,
 \Psi_{lR} , l = e, \mu , \tau  ;
$$
$$
\hspace{1.3cm}i = 1 \hspace{1cm} i = 2 \hspace{1cm} i = 3
$$
$$
\Psi _{iL} = \left(\begin{array}{c} u\\ d \end{array} \right)_L ,
\left(\begin{array}{c} c\\ s \end{array} \right)_L ,
\left(\begin{array}{c} t\\ b \end{array} \right)_L ,
\eqno(2.1)
$$
\par
\noindent
and right singlets of charged leptons and quarks:
$$
\Psi _{iR} = u_{R}, d_{R};\qquad c_{R}, s_{R};\qquad t_{R}, b_{R}
.
$$
\par
The theory is based on the local group $SU(2)\times U(1)$ and
contains two coupling constants $g, g'$. The covariant derivatives
have the following form:
\par
$$
\partial _{\alpha }\Psi _{lL} \rightarrow D_\alpha \Psi_{lL} =
\left[\partial _{\alpha } -
ig {\tau^{i}\over 2} A^{i}_{\alpha }
- ig' \Upsilon ^{lept} B_{\alpha } \right] \Psi _{lL} ,
$$
$$
\partial_{\alpha } \Psi_{iL} \rightarrow  \left[\partial_{\alpha } -
ig {\tau^{i}\over 2} A^{i}_{\alpha }  -
ig' \Upsilon ^{quark} B_{\alpha }\right] \Psi_{iL} ,
\eqno(2.2)
$$
$$
\partial_{\alpha }\Psi _{lR} \rightarrow  \left[\partial _{\alpha } -
i{{g'}\over 2} \Upsilon^{lept} B_{\alpha }\right] \Psi_{lR} ,
$$
$$
\partial_{\alpha }\Psi_{iR} \rightarrow  \left[\partial_{\alpha } -
i{{g'}\over 2} \Upsilon^{quark} B_{\alpha }\right] \Psi_{iR} ,
$$
\par
\noindent
where $A^i_\alpha, B_{\alpha }$ are the gauge fields associated with the
groups $SU(2)_{L}$ and  $U(1)$; $\Upsilon$ is a hypercharge of the leptons
and quarks.
\par
The analog of the Gell-Mann-Nishijima relation in the considered
case is
$$
Q = T^{W}_{3} + {\Upsilon \over 2}  ,
\eqno(2.3)
$$
\par
\noindent
where $Q$ is the electric charge, and $T^{W}_{3}$ is the
third projection of the weak isospin.
\par
For the lepton and quark hypercharges, we obtain the following expressions
using (2.3)
$$
\Upsilon^{lept}_{L} = - 1,\qquad \Upsilon^{quark}_{L} = {1\over 3} ,
\eqno(2.4)
$$
$$
\Upsilon^{lept}_{R} = - 2,\qquad \Upsilon^{quark}_{R} = 2 e_{q} ,
$$
\par
\noindent
where $e_{q}$ is the electrical charge of the corresponding quarks.
\par
Using the standard scheme, we can pass from (2.2) and (2.4) to the
following expression for the interaction Lagrangian:
$$
{\cal L}_{I} = ig j^{K,\alpha } A^{K}_{\alpha}  + ig'\frac{1}{2}
j^{\Upsilon ,\alpha } B_{\alpha } ,
\eqno(2.5)
$$
where
$$
j^{K,\alpha } = \sum^{3}_{i=1} \bar \Psi _{i,L} \gamma ^{\alpha }
{\tau^{K}\over 2} \Psi _{l,L} +
$$
$$
\sum_{l=e,\mu,\tau}  \bar \Psi_{l,L} \gamma ^{\alpha } {\tau^{K}\over 2}
\Psi _{l,L} ,
\eqno(2.6)
$$
and
$$
{1\over 2} j^{\Upsilon,\alpha } = j^{em,\alpha } -
j^{3,\alpha } ,
\eqno(2.7)
$$
$(j^{em,\alpha }$ are electromagnetic current of the quarks and
leptons).
\par
On the transition from the fields $A^{3}_{\alpha }, B_{\alpha }$
to the fields $Z_{\alpha }, A_{\alpha }$
$$
Z_{\alpha } = A^{3}_{\alpha } \cos  \theta _{W} -
B_{\alpha } \sin  \theta _{W} ,
\eqno(2.8)
$$
$$
A_{\alpha } = A^{3}_{\alpha } \sin  \theta _{W} + B_{\alpha } \cos
\theta _{W}  ,
$$
the interaction Lagrangian for the fields $Z_{\alpha }, A_{\alpha
}$   acquires the following form:
$$
{\cal L}^{o}_{I} = i {g\over 2 \cos  \theta _{W}} j^{o,\alpha } Z_{\alpha }
 + ie j^{em,\alpha } A_{\alpha } ,
\eqno(2.9)
$$
where $j^{o,\alpha } = 2 j^{3,\alpha } - 2 \sin ^{2} \theta _{W}
 j^{em,\alpha }$ - neutral current of the standard model.
\par
Note that the Lagrangians (2.5) and (2.9) are obtained for Dirac
(particles) leptons and quarks with charges $g, g'$ or $e, g$
using the principle of local gauge invariance. If $SU(2)_L \times
U(1)$ gauge invariance is required, the masses of all the
particles must be equal to zero (i.e., in this theory particles
cannot have masses [3, 4, 5]. To obtain masses of the particles, the
standard theory of the electroweak interaction based on the
assumption of $SU(2)_L \times U(1)$ gauge invariance is broken
spontaneously down to $U(1)$ through the Higgs mechanism [6]. We
briefly consider this mechanism for three quark family (lepton
masses can be obtained in the analogous manner).
\par
Common remarks concerning the Standard Model are given in the conclusion part
after the discussion of concrete questions.\\
\vspace{1.0cm}
\par
{\bf 2.2. Higgs Mechanism in the standard Model and Remarks}\\

\par
{\bf 2.2.1. Higgs Mechanism in the Standard Model}\\

\par
A doublet of scalar Higgs fields
$$
\Phi  = \left( \begin{array}{c} \Phi ^{(+)}\\ \Phi ^{(o)}\end{array}\right) ,
\eqno(10)
$$
with hypercharge equal to unity (2.3) is introduced. It is assumed
that this doublet interacts with the vector and fermion fields in
such a way that local gauge invariance is not broken. To the
Lagrangian of the electroweak theory there we add the Higgs
potential $V(\Phi ^{+}, \Phi)$
$$
V(\Phi ^{+}, \Phi ) = k(\Phi^{+}, \Phi )^{2} - \mu ^{2}(\Phi^{+}, \Phi ) ,
\eqno(2.11)
$$
($k, \mu^{2}$ are positive constants), which leads to vacuum
degeneracy and to a nonvanishing vacuum expectation value
$<\Phi^{o}>$ of the field $\Phi ^{o}$:
$$
<\Phi ^{o}> = \sqrt{{\mu ^{2}\over 2k}} = \frac{\nu}{\sqrt{2}}
,\qquad \nu = \sqrt{\frac{\mu^2}{k}},
\eqno(2.12)
$$
this means that (fixing the vacuum state) we can generate a mass
term of the fields of the intermediate bosons, fermions, and Higgs
boson.
\par
 In the unitary gauge by using (2.12) we can rewrite $V(\Phi)$ in
the following form ($\nu^2 = \frac{\mu^2}{k}$):
 $$
 V(\Phi) = - \frac{\mu^4}{2} (\nu + \Phi^o)^2 + \frac{k}{4} (\nu + \Phi^o)^4
$$
$$
= -\frac{\mu^4}{4 k} + \mu^2 (\Phi^o)^2 + ... =
-\frac{\mu^4}{4 k} + \frac{m_\Phi^2}{2} (\Phi^o)^2  + ...    ,
\eqno(13)
$$
hence, we see that Higgs boson $\Phi^o$ has mass $m_{\Phi^o}^2 =  2 \mu^2$.
\par
The covariant derivative for Higgs fields is
$$
D_\alpha \Phi = (\partial_\alpha - i g \frac{\tau^i A^i_\alpha}{2} - i
\frac{g'}{2} B_\alpha) \Phi .
\eqno(14)
$$
The kinetic energy term of Higgs bosons (in the unitary gauge) has the
following form:
$$
(D^\alpha \Phi)^{+} D_\alpha \Phi = M^2_W W^{\alpha {+}} W^{-}_\alpha
+ \frac{M^2_Z}{2} Z^\alpha Z_\alpha + ... ,
\eqno(15)
$$
where $W^{\pm}_\alpha = (A^1_\alpha \pm A^2_\alpha)/ \sqrt{2}$, and
their masses are
$$
M^2_W = g^2 \frac{\nu^2}{4},\quad M^2_Z = (g^2 + g'^2)
\frac{\nu^2}{4} .
$$
\par
The quark masses are obtained by using a Lagrangian of Yukawa type
which is $SU(2)_ {L} \times U(1)$ invariant:
$$
{\cal L}_{1} = - \sum^{3}_{i; q=d,s,b}
\bar \Psi_{iL} M^{1}_{iq} q_{R} \bar \Phi + H. C.  ,
\eqno(2.16)
$$
$$
{\cal L}_{2} = -  \sum^{3}_{i; q=u,c,t}
\bar \Psi_{iL} M^{2}_{iq} q_{R} \bar \Phi + H. C. ,
$$
where $M^{1}, M^{2}$ - complex $3 \times 3$ matrix, and $\bar \Phi$
$$
\bar \Phi = i\tau_{2} \Phi ^{*} = \left(\begin{array}{c} \Phi^{o*} \\
-\Phi^{+*} \end{array} \right) ,
\eqno(2.17)
$$
is a doublet of Higgs fields with hypercharge $Y = -1$.
\par
Taking into account (2.12) and using the gauge invariance of the
Lagrangian (2.13), (2.17), we can choose (in the unitary gauge)
$$
\Phi (x) = \left(\begin{array}{c} 0\\ {{\nu  + \Phi^{o}(x)}\over \sqrt{2}}
\end{array} \right) ,\qquad
\bar\Phi (x) = \left(\begin{array}{c}{{\nu  + \Phi^{o}(x)}\over \sqrt{2}}\\0
\end{array}\right) ,
\eqno(2.18)
$$
where $\Phi^{o}(x)$ is the neutral scalar Higgs field.
\par
Substituting (2.18) in (2.16), for the quark masses we obtain
the expressions
$$
{\cal L}_{1} = - \bar p_{L} {M'}_{1} p_{R} + H. C. ,
\eqno(2.19)
$$
$$
{\cal L}_{2} = - \bar n_{L} {M'}_{2} n_{R}  + H. C. ,
$$
where
$$
p_{L,R} = \left(\begin{array}{c} u_{L,R}\\ c_{L,R}\\
t_{L,R}\end{array}\right) ,\qquad
n_{L,R} = \left(\begin{array}{c} d_{L,R}\\ s_{L,R}\\ b_{L,R}
\end{array}\right)  .
$$
Thus, the elements ${M'}_{1}, {M'}_{2}$ of the quark mass matrix
are equal to the constants of the quark-Higgs-boson Yukawa
coupling up to the factor $ \nu $.\\

{\bf 2.2.2. Remarks to the Higgs Mechanism in the Electroweak Model}\\

\par
We know that quarks, leptons and vector bosons have their  masses
in every point of the Universe. Then Higgs fields must fill the
Universe  and since the masses are real masses, then Higgs fields
must also be real (here we have an analogy with the
superconductivity). If Higgs field is real, then the energy
density of this field  is $\rho_{Higgs} \sim 2\cdot 10^{49} GeV/cm^{3}$ [6, 7]
(see also references in [7]).
It is a huge value. The measured energy density in the Universe is
$\rho_{Univ} \sim  10^{-4} GeV/cm^{3}$. Then the relation of the energy
density of the Higgs fields to the measured energy value is
$$
\rho_{Higgs}/ \rho_{Univ} \sim 10^{53} .
\eqno(2.20)
$$
It is obvious that the Higgs mechanism must be excluded.
Besides, Higgs  mechanism contains a contradiction [8].
We see that the Higgs
mechanism is used to construct a theory without singularity,
however our object is to construct a realistic physical theory.
Then arises question: what are mass sources? There is some arguments
that mass sources must be a mechanism which is analogous to the
strong interactions [9], i.e. interactions between
subparticles of quarks  and leptons, then the problem of
singularity of the theory does  not arise. \\

\newpage
\par
{\bf 2.3. Running Couple Constant in the Standard Weak Interactions and
Remarks}\\

{\bf 2.3.1. Running Couple Constant in the Standard Weak Interactions}\\

\par
It is supposed that in the electroweak model  the couple constants
$g, g'$ depend on transfer momenta [10] and the equation for $g(Q^2)$ has
the following form:
$$
\frac{d g^{-1}}{d ln Q^2} = \frac{1}{4 \pi} \left[ \frac{22}{3} -
\frac{4 F}{3} \right] ,
\eqno(2.21)
$$
where $F$ is family numbers ($F = 3$) and  $Q^2$ is a transfer
momentum squared (here we consider only a weak part of the
electroweak model since in the electromagnetic interactions there
is renormalization of the couple constant). It means that in the
weak interactions the vacuum polarization takes place as in the
strong and electromagnetic interactions. It is necessary to
remember that in the weak interactions in contrast to these
interactions only the left components of fermion participate in
the weak interactions.
\par
If the couple constant of the weak interaction  is renormalized,
then the effective masses of fermions in matter also will be
changed, i.e. the standard weak interaction can generate effective
masses. It means that the resonance enhancement of neutrino oscillations
in matter [11] will take place at the weak interactions. \\

\par
{\bf 2.3.2. Remarks About the Couple Constant of the Standard Weak
Interactions}\\

\par
As we have stressed above, the distinctive feature of the weak
interactions is violation of $P$ parity in these interactions. Now
we will consider consequences of the distinctive feature for the couple
constant of the weak interactions.

\par
The simplest method to prove the absence of  polarization in
vacuum and matter is:
\par
   If we put an electrical (or strong) charged particle in  the vacuum,
there arises polarization of vacuum. Since the field around the
particle is spherically symmetrical, the polarization must also be
spherically symmetrical. Then the particle will be left at rest
and the law of energy and momentum conservation is fulfilled.
\par
If we put a weakly interacting particle (a neutrino) in the vacuum,
then, since the field around the particle has a left-right
asymmetry (weak interactions are left interactions with respect to
the spin direction), polarization of the vacuum must be
nonsymmetrical, i.e. on the left side there arises maximal
polarization and on the right-there is zero polarization. Since
polarization of the vacuum is asymmetrical, there arises
asymmetrical interaction of the particle (the neutrino) with
vacuum and the particle cannot be at rest and will be accelerated.
Then neutrino will get the energy-momentum from the vacuum and
the law of energy momentum conservation will
be violated. The only way to fulfil the law of energy-momentum
conservation is to require that polarization of vacuum be absent at
the weak interactions. The same situation will take place in
matter (do not mix it up with particle acceleration at the weak
interactions!) .
\par
About a direct method for proving the absence of polarization in the weak
interactions see in Section 2.5.
\par
It is interesting to remark that in the gravitational interaction
the polarization does not exist either [12]. \\

{\bf 2.4. Generation of Masses in the Standard Model
and the Mechanism of Resonance Enhancement of Neutrino
Oscillations in Matter and Remarks}\\

\par
{\bf 2.4.1. Generation of Masses in the Standard Model and
the Mechanism of Resonance Enhancement of Neutrino
Oscillations in Matter}\\

At present there is a number of papers published (see [13] and
references there) where by using the Green's function method it is
shown that the weak interactions can generate the resonance
enhancement of neutrino oscillations in matter (it means that the
weak interaction can generate masses, i.e.
the energy $W$ of matter polarization by neutrinos (or the energy of the
matter response ) will be differ from zero $W \not= 0$ [5, 14]).
This result is a
consequence of using the weak interaction term $H^{int}_\mu =
V_\mu \frac{1}{2}(1- \gamma_5)$ in an incorrect manner, and in the
result they have obtained that the right-handed components of the
fermions participate in the weak interactions [15, 16].\\
\vspace{1cm}

{\bf 2.4.2. Remarks to the Problem of Generation of Masses in the Standard
Model and the Mechanism of Resonance Enhancement of Neutrino
Oscillations in Matter}\\

\par
In three different approaches: by using mass Lagrangian [3, 17], by
using the Dirac equation [4, 17], and using the operator formalism
[5, 14], I considered the problem of the mass generation in the
standard weak interactions. The result was: the standard weak
interaction cannot generate masses of fermions since the
right-handed components of fermions do not participate in these
interactions. Then using this result in works [5, 14] it has been
shown that the effect of resonance enhancement of neutrino
oscillations in matter cannot exist (existence of this effect means
that the law of energy-momentum conservation is violated).
\par
The experimental data on energy spectrum and day-night effect obtained in
Super-Kamiokande [18] (energy spectrum of neutrinos is not distorted,
day-night effect is within the experimental mistakes) and the results
obtained in SNO [19] have not confirmed this effect. Besides, this effect can
be realized only at the violation of the law of the energy-momentum
conservation (see Ref. [20]).\\

\newpage
{\bf 2.5. Problem of Connected (Bounded) States in the Standard Model
of Weak Interactions}\\

\par
Now consider the problem of eigenstates and eigenvalues in the
weak interactions [16]. Let $\hat F$ be an operator and we divide
it into two parts. The first part $\hat A$ characterizes the free
particle, and the second part $\hat B$ is responsible for the weak
interaction, then
$$
\hat F = \hat A + \hat B ,
$$
$$
\hat F\Psi = \hat A\Psi + \hat B\Psi ,
\eqno(2.22)
$$
and the mean value of $\hat F$ is
$$
(\bar\Psi, \hat F\Psi) = (\bar\Psi, \hat A\Psi) + (\bar\Psi_R,
\hat B\Psi_L) + (\bar\Psi_L, \hat B\Psi_R) =
$$
$$
(\bar\Psi, \hat A\Psi) + (\bar \Psi_R \equiv 0)(\bar\Psi_R, \hat
B\Psi_L) +
\eqno(2.23)
$$
$$
+(\Psi_R \equiv 0)(\bar\Psi_L, \hat B\Psi_R) =
(\bar\Psi, \hat A\Psi) .
$$
The obtained result means that in the weak interactions there
cannot arise the connected states in contrast to the strong and
electromagnetic interactions.
\par
Besides, the average of the polarization operators  is equal to
zero, i.e. the polarization of the matter is absent. In the same
way we can show that the equation for renormcharge for the weak
interaction is equivalent to the equation for the free charge,
i.e. renormcharge $g(t)$ in the weak interactions (where
$t = Q^2$ is a transfer momentum squared) does not change and $g(t) =
const$ [21, 16] in contrast to renormcharges $\alpha(t), g_{str}(t)$ of the
electromagnetic and strong interactions [1] (it is necessary to
remark that the neutral current of the weak interactions includes
a left-right symmetrical part which is renormalized).
\par
So, at the weak interaction the connected states cannot arise, but
also as in the strong and electromagnetic interactions, in the weak
interactions we can use the perturbative theory but in this case
the propagators must be the propagators of free particles (without
renormalization).

\section{Conclusion}

So, we have considered some remarks to the standard approach to the weak
interactions. Higgs mechanism contains contradictions, therefore cannot
be considered as a realistic mechanism of mass generations. The couple
constant of the weak interactions cannot be changed in dependence on momenta
transfer (it leads to violation of the law of energy-momenta conservation),
i.e. there it must be constant. It means that there is no
charge vacuum polarization or renormalization of the couple constant
of the weak interactions.  Then it is clear that the resonance
enhancement of neutrino oscillations in matter cannot exist either.
Masses are not generated and connected states cannot form in
these interactions since $P$-parity is violated.  Then also as in the strong
and electromagnetic interactions, in the weak interactions we can use
the perturbative theory but in this case the propagators must be the
propagators of free particles without renormalization.  Probably the
standard weak interactions model with $W, Z $ exchanges is analogous to
the strong interactions with $\pi$ or $\rho$ exchanges and it is
necessary to find fundamental interactions which generate this picture
of the weak interactions and generate masses.
\par
It is also possible that in the weak interactions the couple constant
renormalization absent for singularity compensations as it take place in
the supersymmetric theories [22]. \\

\par
{\bf References}
\par
\noindent
1. Review of Part. Prop., The European Phys. Journ. C,
\par
2000, v.15, N1-4.
\par
\noindent
2. Glashow S.L.  Nucl. Phys. 1961, vol.22, p.579 ;
\par
Weinberg S., Phys.  Rev. Lett., 1967, vol.19, p.1264 ;
\par
Salam A., Proc. of the 8-th    Nobel  Symp.,  edited  by     N.
Svarthholm
\par
(Almgvist and Wiksell,  Stockholm) 1968,p.367.
\par
\noindent
3. Beshtoev Kh.M., JINR  Commun. E2-93-44, Dubna, 1993.
\par
\noindent
4. Beshtoev Kh.M., JINR  Commun. E2-93-167, Dubna, 1993.
\par
\noindent
5. Beshtoev Kh. M., Hadronic Journ. 1999, vol.22, p.477.
\par
\noindent
6. Higgs P.W., Phys. Lett., 1964, Vol.12, p.132;
Phys.Rev., 1966,
\par
vol.145, p.1156;
\par
Englert F., Brout R., Phys. Rev. Lett, 1964, vol.13,p.321;
\par
Guralnik G.S. et al.,  Phys.  Rew. Lett, 1964, vol.13,p.585.
\par
\noindent
7. Langacker p., Phys. Rep. 1981, v.72, N4.
\par
\noindent
8. Beshtoev Kh. M. and Toth L., JINR Commun. E2-90-475,
\par
Dubna, 1990;
\par
Beshtoev Kh. M., JINR Commun. E2-92-195, Dubna,1992.
\par
\noindent
9. Beshtoev Kh. M., JINR Commun. E2-99-137, Dubna,1999;
\par
JINR Commun. E2-99-306, Dubna,1999.
\par
\noindent
10. D.J. Gross and F. Wilczec,  Phys. Rev. D9, 1974, p.980;
\par
H.D. Politzer, Phys. Rev. Lett.,  30, 1973, p.1346;
\par
A. Zee, Phys. Rev. D7, 1973, 3630;
\par
G. Kane, Modern Elementary Particle Physics,
\par
Univ. Michigan, 1987.
\par
\noindent
11. Wolfenstein L., Phys. Rev.D, 1978, vol.17, p.2369;
\par
Mikheyev S.P., Smirnov A.Ju., Nuovo Cimento, 1986, vol.9,p.17.
\par
\noindent
12. Beshtoev Kh. M., JINR Commun. P2-2000-137, Dubna,2000;
\par
Phys. Essays, 2000, v.13, p.593.
\par
\noindent
13. C.Y. Cardall, D.J. Chung, Phys. Rev.D 1999, v.60, p.073012.
\par
\noindent
14. Kh. M. Beshtoev, HEP-PH/9912532, 1999;
\par
Hadronic Journal, 1999, v.22, p.235.
\par
\noindent
\par
\noindent
15. Beshtoev Kh. M., JINR Commun. E2-2000-30, Dubna, 2000.
\par
\noindent
16. Beshtoev Kh. M., JINR Commun. E2-2001-65, Dubna, 2001.
\par
\noindent
17. Kh.M. Beshtoev, Fiz. Elem. Chastitz At. Yadra (Phys.
\par
Part. Nucl.), 1996, v.27, p.53.
\par
\noindent
18. Kajita T., Report on Intern. Conf. "Neutrino98" Japan,
\par
June, 1998;
\par
Fukuda S. et al., Phys. Rev. Lett., 2001, v.86, p.5651.
\par
\noindent
19. Ahmad Q. R. et al., Internet Pub. nucl-ex/0106015, June 2001.
\par
\noindent
20. Kh. M. Beshtoev, JINR Commun. D2-2001-292, Dubna, 2001.
\par
\noindent
21. Kh. M. Beshtoev, JINR Commun. E2-94-221, Dubna, 1994.
\par
\noindent
22. P. Howe et al., Nucl. Phys. B214, 1983, p.519;
\par
S. Mandelstam, Nucl. Phys. B213, 1983, p.149;
\par
P. Howe et al., Phys. Lett. 124B, 1983, p.55;
\par
A. V. Ermushin et al., Nucl. Phys. B281, 1987, p.72.

\end{document}